\documentclass[12pt]{article}

\usepackage{isolatin1}
\usepackage{colordvi}
\usepackage{color}
\usepackage{amsmath}
\usepackage{amssymb}
\usepackage{latexsym}
\usepackage{theorem}
\usepackage{epic}
\usepackage{eepic}
\usepackage{epsfig}

\newcommand{\lstmarksep}{\hspace{1em}}

\newcommand{\markendlst}[1]{
  \begin{flushright}
    {\bf #1\lstmarksep}$\Box$
  \end{flushright}
}

\theoremstyle{break}
\theorembodyfont{\rmfamily}

\newtheorem{example}{Example}

\newtheorem{definition}{Definition}

\newcommand{\incr}{\mathord{\uparrow}}

\newcommand{\derives}{\qquad\Longrightarrow\qquad}

\newcommand{\Scomb}{\mathord{\mathbf{S}}}
\newcommand{\Kcomb}{\mathord{\mathbf{K}}}

\title{The Sources of Certainty\\ in Computation and Formal Systems}

\author{Michael J.\ O'Donnell\\ \textit{The University of Chicago}}

\date{2 November, 1999\\ Bibliography corrected 17 November, 1999}

\begin{document}

\maketitle

\begin{abstract}
In his \emph{Discourse on the Method of Rightly Conducting the Reason,
and Seeking Truth in the Sciences}, Ren\'e Descartes sought ``clear
and certain knowledge of all that is useful in life.'' Almost three
centuries later, in ``The foundations of mathematics,'' David Hilbert
tried to ``recast mathematical definitions and inferences in such a
way that they are unshakable.'' Hilbert's program relied explicitly on
\emph{formal systems} (equivalently, \emph{computational} systems) to
provide certainty in mathematics. The concepts of computation and
formal system were not defined in his time, but Descartes' method may
be understood as seeking certainty in essentially the same way.

In this article, I explain formal systems as concrete artifacts, and
investigate the way in which they provide a high level of
certainty---arguably the highest level achievable by rational
discourse. The rich understanding of formal systems achieved by
mathematical logic and computer science in this century illuminates
the nature of programs, such as Descartes' and Hilbert's, that seek
certainty through rigorous analysis.
\end{abstract}

I presented this paper on 30 October 1999, at the 1999--2000 Sawyer
Seminar at the University of Chicago, \emph{Computer Science as a
Human Science: the Cultural Impace of Computerization}.

\section{Scholarly Disclaimer}
I use the writings of several excellent thinkers---Descartes, Hilbert,
Curry, Mac Lane---to stimulate insight regarding the deliberate and
accidental use of formal systems in philosophical and mathematical
thought. I believe that all of my quotes are interpreted sensibly, but
I am \emph{not} trying to infer the actual beliefs or intentions of
the authors. Nor am I trying to make a fair representation of the
whole of their works. Rather, I am taking the writings of these
authors as tools to be applied at need, and selecting only those
writings that relate to my topic.

The modern authors in my list are all well versed in the concept of
formal system, and it is clear that they intended to discuss such
systems, whether or not they would agree with my own use of those
discussions. Descartes had no access to a definition of formal system
(although he focused much attention on Euclidean geometry, which we
recognize now as a formal system). I will argue that it makes sense,
in retrospect, to apply many of Descartes' ideas to formal systems,
but I make no claim that Descartes himself would have done so under
some sort of counterfactual assumptions.

My topic, then, is the illumination of certain lines of thought using
the concept of formal systems. I welcome the contributions of great
thinkers to that topic, whether those contributions are intentional or
accidental.

\section{Computational Concepts\\ in Uncomputational Topics}
This article constitutes part of a larger program of my own, to
connect computer science to other disciplines in an unusual way. I am
familiar with interdisciplinary work
\begin{itemize}
\item that uses computations on electronic devices to calculate the
consequences of various theories, or
\item that applies the methods of other disciplines to study the
behavior of systems comprising computers and other entities (such as
people).
\end{itemize}
I intend a third form of connection between computer science and other
disciplines
\begin{itemize}
\item that applies \emph{concepts} from computer science to illuminate
phenomena studied by other disciplines.
\end{itemize}

Computation is a sort of behavior, and the electronical gizmos that we
now call ``computers'' are not the only devices that exhibit that sort
of behavior. A few decades ago, ``computer'' was a job title for a
person, and Alan Turing called his proposed gizmo an ``Automated
Computing Engine,'' to distinguish it from a human ``computer.'' Other
systems in the world may exhibit computational behavior
unconsciously. The concept of computation was developed quite clearly
before the construction of automatic computers. But, familiarity with
the behavior of automatic computers makes some ideas, that once seemed
highly arcane and subtle, intuitively accessible to a much wider range 
of thinkers than before.

Computational concepts may now be applied to understand phenomena in
quantum physics, thermodynamics, probability, genetics, and
immunology. In this article, I follow another such application to
philosophy.

\section{Formal Systems and Computation}
The phrase \emph{``formal system''} is widely misunderstood, even by
mathematicians who profess formalist foundations for their
work. Perhaps a quick way to refine the understanding of ``formal'' is
to consider its opposite. In common use, the opposite of ``formal''
might be ``casual,'' or ``relaxed,'' or ``unrigorous.'' Mathematicians
often call a derivation ``informal'' if it is incomplete or not
described quite thoroughly. None of these is a sensible opposite to
``formal'' for our purposes. A formal system is one that deals with
the forms of arrays of symbols and relations between them---a proper
opposite is a \emph{contentual} system that deals with the content or
meaning of arrays of symbols.

Here is an example description of a formal system capturing a tiny bit
of mathematics.
\begin{example}[A formal system for incrementing integers.]
\label{exa:incr}
Figure~\ref{fig:incrsys} describes derivation rules for a simple
formal system.
\begin{figure}
\newcommand{\commwid}{2.5in}
\newcommand{\intspace}{2ex}
\begin{xalignat*}{2}
& \derives x=x & & \text{%
\begin{minipage}[t]{\commwid}
You may start with two copies of the same sequence of `$0$'s
`$1$'s and `$\incr$'s, with `=' between them.
\end{minipage}%
} \\[\intspace]
0\incr & \derives 1 & & \text{%
\begin{minipage}[t]{\commwid}
You may replace the pair `$0$' followed immediately by `$\incr$' with
`$1$'.
\end{minipage}%
} \\[\intspace]
1\incr & \derives \incr 0 & & \text{%
\begin{minipage}[t]{\commwid}
You may replace the pair `$1$' followed immediately by `$\incr$' with
`$\incr 0$'.
\end{minipage}%
} \\[\intspace]
=\incr & \derives =1 & & \text{%
\begin{minipage}[t]{\commwid}
You may replace `$\incr$' by `$1$' when it follows immediately after `$=$'.
\end{minipage}%
}
\end{xalignat*}
\caption{\label{fig:incrsys} Derivation rules for a formal system for
incrementing integers}
\end{figure}
The left column presents the rules schematically, and the right column
provides an alternate presentation in English. In the formal system
described above, we may derive `$11\incr=100$' as follows:
\begin{align*}
11\incr & =11\incr \\
11\incr & =1\incr 0 \\
11\incr & =\incr 00 \\
11\incr & =100
\end{align*}
\markendlst{End Example~\ref{exa:incr}}
\end{example}
The formal system of Example~\ref{exa:incr} may be understood as a
presentation of the theory of nonnegative integers expressed in binary
notation, with the operation of adding one indicated by placing
`$\incr$' to the right of an integer. The derivation in the example
may be understood to demonstrate, in more familiar notation, that
$3+1=4$.

One more example suggests the variability and power of formal systems, 
at the cost of appearing very esoteric in interpretation.
\begin{example}[A simple formal system capable of all computation.]
\label{exa:SK}
Figure~\ref{fig:SK} shows the two rules for manipulation in a famous
formal system called the \emph{Combinator Calculus}.

\begin{figure}
\epsfig{file=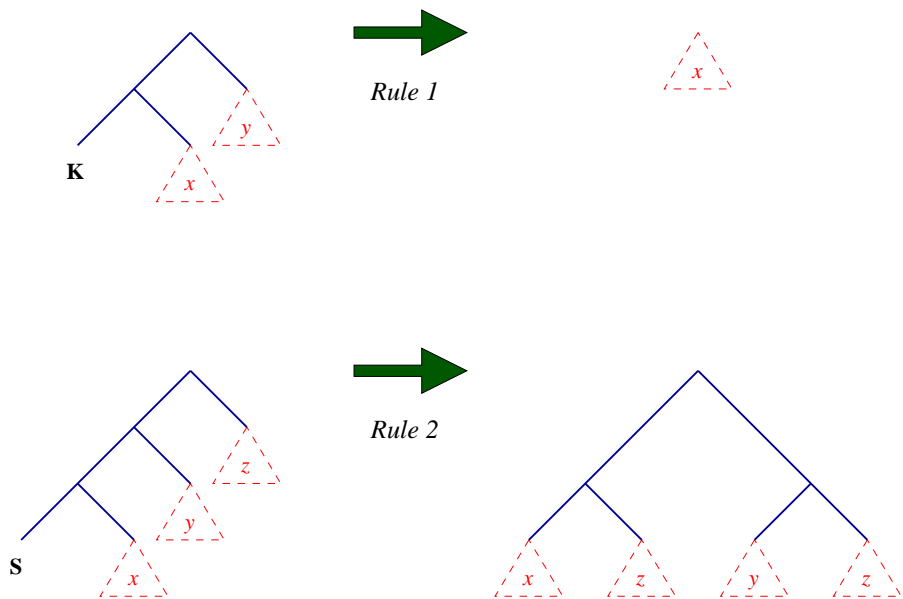}
\caption{\label{fig:SK} Derivation rules for the Combinator Calculus}
\end{figure}
The English description of the rules is too long and tangled to be
worth inspecting here. The pictures should be clear enough, as long as
we understand that
\begin{itemize}
\item The system deals entirely with finite binary branching tree
diagrams, where the end of each path is labelled with exactly one of
the symbols `$\Scomb$' or `$\Kcomb$'. Such a tree diagram is called a
\emph{combinator}.
\item You may start with any combinator.
\item In Figure~\ref{fig:SK}, the $x$, $y$, and $z$ in dashed
triangles may be replaced by any combinators, as long as in each
application of a rule, each of the $x$ triangles is replaced by a copy 
of the same combinator, similarly for each of the $y$ triangles and
each of the $z$ triangles.
\item When a structure of the form given by the left-hand side of one
of the two rules in Figure~\ref{fig:SK} appears anywhere within a
combinator, you may replace that structure by the corresponding
structure on the right-hand side of the same rule.
\end{itemize}
In the formal system of the Combinator Calculus, we may replace a
certain combination of four `$\Kcomb$'s and two `$\Scomb$'s by the
combination of the two `$\Scomb$'s, using the derivation in
Figure~\ref{fig:SKsec}.
\begin{figure}
\begin{center}
\epsfig{file=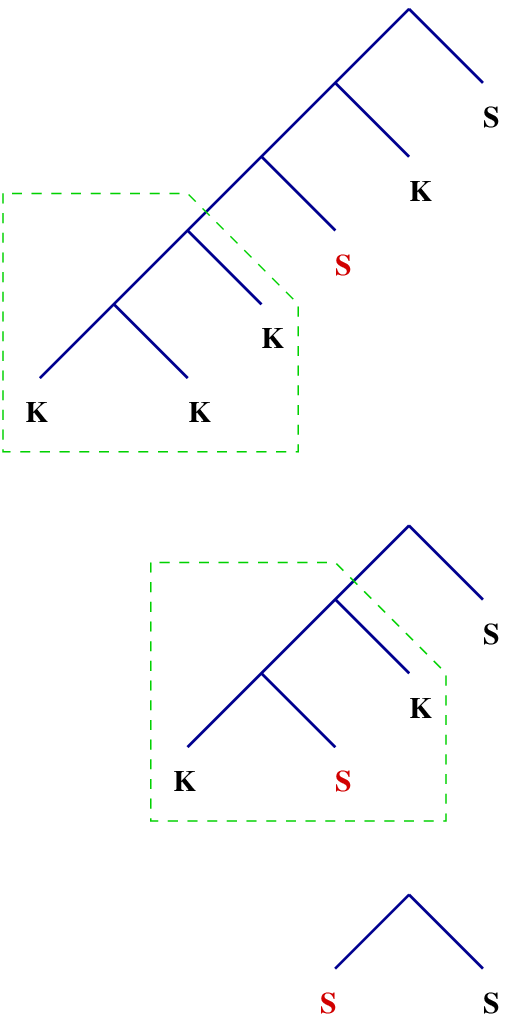}
\end{center}
This derivation takes two steps with \emph{Rule 1} to derive a tree
with two $\Scomb$s. Intuitively, the leftmost two $\Kcomb$s select the
red $\Scomb$ from the surrounding $\Kcomb$s.

The five-sided dashed figures are not part of the derivation: they
just show where the rules are applied.
\caption{\label{fig:SKsec} A derivation in the Combinator Calculus}
\end{figure}
We may always eliminate a certain combination of an `$\Scomb$' and two
`$\Kcomb$'s, as shown in the schematic derivation of
Figure~\ref{fig:SKident}.
\begin{figure}
\begin{center}
\epsfig{file=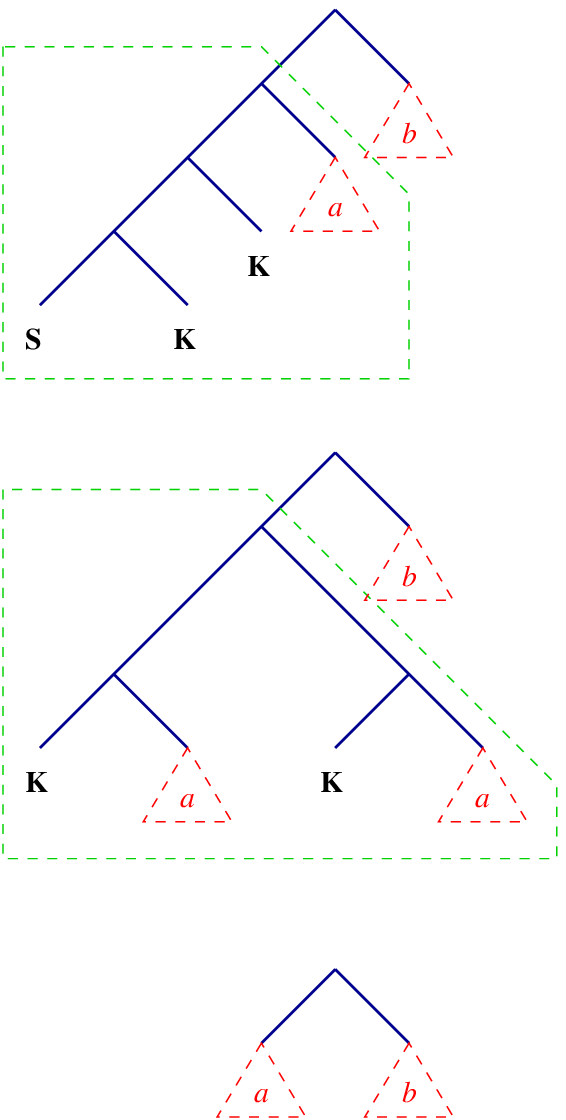}
\end{center}
This schematic derivation takes two steps, first with \emph{Rule 2}
and then with \emph{Rule 1}, to reach the schematic tree at the
bottom. Intuitively, the combination of one $\Scomb$ and two $\Kcomb$s 
above acts as an \emph{identity operator} applied to the tree that
fills in for $a$.

The five-sided dashed figures are not part of the derivation: they
just show where the rules are applied.
\caption{\label{fig:SKident} A schematic derivation in the Combinator
Calculus}
\end{figure}
\markendlst{End Example~\ref{exa:SK}}
\end{example}
The useful interpretations of the Combinator Calculus are too long to
describe here. The interesting qualities of this formal system for our 
purposes are
\begin{itemize}
\item it is best described in terms of binary branching tree diagrams, 
instead of sequences of symbols;
\item although it has only two rather simple rules, it is capable of
deriving values of every function that is computable by every other
formal system.
\end{itemize}
The derivation in Figure~\ref{fig:SKsec} shows how two `$\Kcomb$'s may
be used to select the middle of a sequence of three structures (in
this case, the middle structure is just an `$\Scomb$', marked red to
help you follow the process). Notice that the regions in the green
dashed lines contain the pattern in the left-hand side of \emph{Rule
1}.
\begin{enumerate}
\item In the first step, the leftmost $\Kcomb$ matches the $\Kcomb$ in
the left-hand side of \emph{Rule 1}, and the next two $\Kcomb$s to the 
right fill in for the $x$ and $y$. Following \emph{Rule 1}, this whole 
structure with three letters is replaced by $\Kcomb$ (filling in for
$x$ in the rule).
\item In the second step, the leftmost $\Kcomb$ matches the $\Kcomb$
in the left-hand side of \emph{Rule 1}, the red $\Scomb$ fills in for
the $x$, and the second $\Kcomb$ fills in for the $y$. Following
\emph{Rule 1}, this whole structure with three letters is replaced by
$\Scomb$ (filling in for $x$ in the rule).
\end{enumerate}

We quickly get bored doing one derivation at a time, and notice
\emph{schematic derivations}---patterns that can be expanded into
infinitely many different but similar derivations.  The schematic
derivation in Figure~\ref{fig:SKident} shows how an appropriate
combination of an `$\Scomb$' and two `$\Kcomb$'s represents the
identity function: when applied to any combinator $a$ it produces
(after two derivation steps) $a$ alone. Figure~\ref{fig:SKident} is
\emph{schematic} in the sense that, although it does not present a
derivation per se, every systematic replacement of the triangles
containing $a$ and $b$ in the figure yields a derivation.
\begin{enumerate}
\item In the first step, the leftmost $\Scomb$ matches the $\Scomb$ in
the left-hand side of \emph{Rule 2}, the two $\Kcomb$s fill in for $x$ 
and $y$, and whatever fills in for $a$ also fills in for $z$ in the
rule. Following \emph{Rule 2}, this whole structure is replaced by the 
combination of two $\Kcomb$s and two $a$s.
\item In the second step, the leftmost $\Kcomb$ matches the $\Kcomb$
in the left-hand side of \emph{Rule 2}, the leftmost $a$ fills in for
$x$, and the second combination of $\Kcomb$ with $a$ fills in for
$y$. This whole structure is replaced by a single copy of $a$.
\end{enumerate}

The idea of a schematic derivation is worth some attention, as it
illustrates the highly reflexive way in which formal systems provide
reasoning power. Most of the intuitively important observations about
formal systems are schematic---they are observations of
\emph{patterns} in the derivations of the formal system, rather than
individual derivations. But, there is another formal system containing
the derivations of the Combinator Calculus, and also derivations with
formal variable symbols. Individual derivations in the Combinator
Calculus with variables correspond to schematic patterns of
derivations in the Combinator Calculus, in a rigorous way. There is
yet another formal system that models the correspondence between
schematic derivations in the Combinator Calculus and derivations in
the Combinator Calculus with variables.

\begin{figure}
\begin{center}
\epsfig{file=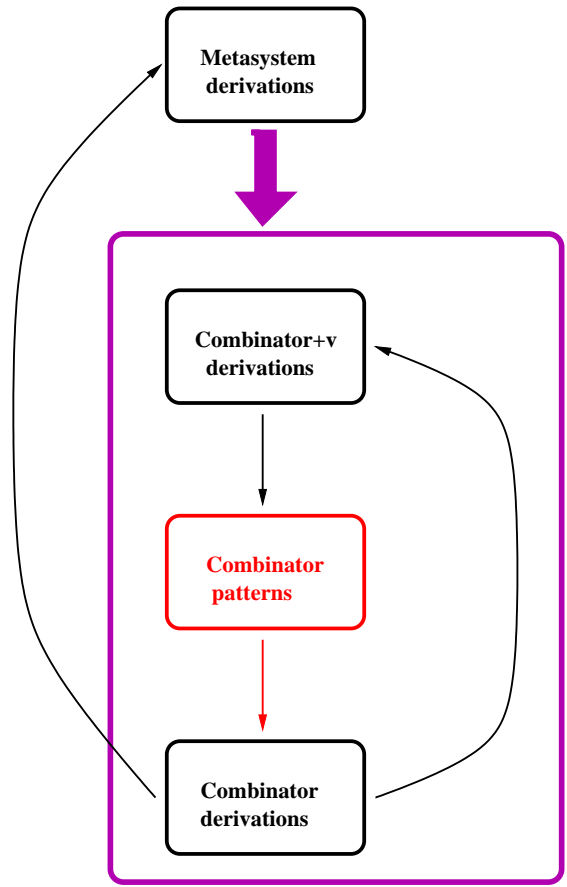}
\end{center}
\caption{\label{fig:modeling} Variations on Combinator Calculus and
their modeling relations}
\end{figure}

But, the trickiest twists are yet to come. The Combinator Calculus was
designed specifically to be able to simulate the behavior of systems
with variables, in a variable-free style. So, the Combinator Calculus
contains a precise model of the behavior of the Combinator Calculus
with variables, and therefore single derivations in the Combinator
Calculus can demonstrate the behaviors of schematic derivations in the
Combinator Calculus. And, there's a formal system that models the
correspondence between the Combinator Calculi with and without
variables, and the Combinator Calculus contains a model of that
system, and \dots. Figure~\ref{fig:modeling} suggests the systems and
relations described above, but of course the real picture is
infinitely large, and infinitely more complicated.

Formal systems can be used to study one another in highly tangled and
reflexive, but also powerful and productive, ways. This tangled
reflexivity has a lot to do with the effectiveness of formal methods,
and it is no doubt the source of a lot of the confusion about the
precise relationship between formal systems, mathematics, and the rest
of the world. Saunders Mac Lane traces the reflexivity of formal
systems within mathematics rather thoroughly in
\emph{Mathematics, form and function}.

Those who pay attention to such things appear to accept unanimously
that
\begin{itemize}
\item formal systems, such as the one described in Example~\ref{exa:incr},
may be treated as pure symbolic games, without referring to any
meaning that might be associated with their symbols; and
\item formal systems are at the heart of mathematics.
\end{itemize}
Unfortunately, many thinkers (including working mathematicians) also appear to
misconstrue the detailed nature of formal systems, and their
particular relation to mathematics.

\paragraph{Symbols and form as physical objects, vs.\ abstract constructs.}
A natural and common view of formal systems holds that they describe
physical operations that may be applied to ink on paper, or at least
to specifically typographical presentations of symbols with specific
geometrical shapes. This view makes formal systems satisfyingly
concrete and physical at first glance, but it also makes them very
puzzling, since the particular qualities of, for example, ink spread
on paper, have little to do with the interesting qualities of formal
systems. The actual practice of presenting formal systems argues
against a specific physical view. Practitioners of formal studies
routinely accept and appear to understand presentations with ink on
paper, chalk on slate, electron beams bombarding phosphors, sound
vibrations in the air, and electrochemical systems in the brain. The
mental presentation is uniquely important, since it alone is
irreplacable when we use formal systems as thinking tools. But, in
essence, it is still just another presentation.

Formal systems surely have to do with the forms of arrays of symbols,
rather than the meanings of such arrays. But, the symbols and arrays
are best understood as mental constructs, which may be communicated
through any physical presentation that satisfies all the parties to a
discussion. Haskell Curry explains this view rather well in
\emph{Outline of a Formalist Philosophy of Mathematics}. Curry even
challenges the common notion that arrays of symbols should always be
presented as linear sequences of characters, a widely accepted view
that appears to me to derive accidentally from typographical
technology. He argues that graphical structures, usually in the form
of trees such as those in Example~\ref{exa:SK}, are more suitable for
most applications of formal systems in mathematics. In principle,
formal systems should be defined using whatever sorts of layouts of
symbols are most effective for their particular uses.

\paragraph{Mathematics as a formal system, vs.\ mathematics as a study 
of formal systems.}
The second misconstrual of formal systems has to do with their
particular relation to mathematics. Many thinkers, especially working
mathematicians, appear to accept the view that mathematics is a game
played out by the rules of some formal system, or perhaps a collection
of such games. That is, they accept the notion that mathematicians at
work are carrying out derivations in a formal system. A variation on
this view holds that, while mathematicians at work may employ mental
steps that do not follow the rules of any well identified formal
system, the correctness of that work depends on the possibility of
recasting it as such a formal derivation. The actual practice of
mathematics is viewed as a slightly risky, but more efficient,
prediction of the results of particular formal derivations. Mac Lane
calls the notion of mathematics as an arbitrary formal game ``vulgar
formalism.'' I cannot find an explicit profession of ``vulgar
formalism,'' even with a more dignified title, in print. But, I have
heard mathematicians and users of mathematics express such views in
casual conversation, and they sounded serious.

If mathematics is indeed the playing out of a game whose rules have
only to do with the superficial forms of symbols, and nothing to do
with their useful meanings, we are naturally puzzled why mathematics
appears to be useful as a tool for practical work. R.~W. Hamming and
E.~P. Wigner expressed this puzzlement in papers called ``The
unreasonable effectiveness of mathematics'' and ``The unreasonable
effectiveness of mathematics in the natural sciences,'' respectively,
whose titles are quoted widely by mathematicians who worry that a mere
formal system should not capture scientific ideas with real-world
content. I think that it is much more sensible to view mathematics as
a study with real objective content, whose effectiveness is a result
of the widespread applicability of that content. (Hamming and Wigner
do not seem to hold a formalist view of mathematics themselves, and
their papers contain a lot of interesting discussion of the actual
practice of mathematics and science.)

Rather than mathematics itself being a formal game, I support the view
that mathematics is essentially a rigorous study of the behavior of
formal systems. Confusion about the relation of mathematics to formal
systems probably derives largely from the way in which mathematical
tools are particularly effective for studying technical methodological
issues in the practice of mathematics itself. For example, in
mathematical logic we study formal systems that are designed to
illuminate mathematical reasoning. It is not hard to misconstrue this
application of formal tools to the study of mathematics as an a priori
embedding of mathematics into a formal system.

Saunders Mac Lane explains the role of formal systems in mathematics
very thoroughly in \emph{Mathematics, Form and Function}, including
the convolutions induced by the reflexive uses of mathematics. Haskell
Curry covered similar ground earlier in
\emph{Outline of a Formalist Philosophy of Mathematics}. I find
Curry's discussion less satisfying in its treatment of actual
mathematical activity, but he contributes a slogan which I find
extremely useful as long as it is used to stimulate thought, rather
than as a final conclusion:
\begin{quote}
Mathematics is the science of formal systems.
\end{quote}
That is, mathematics is not just the playing of a formally defined
game; it has objective content---the qualities of formal systems. I
propose another slogan to stimulate similar thought from a slightly
different point of view:
\begin{quote}
The content of mathematics is form.
\end{quote}
For present purposes, it is important that formal systems are real,
objective, but not physical, things, that we can be highly certain
about their behavior, and that the objective study of formal systems
is at least a large part of the content of mathematics. But, it is not
important to maintain Curry's slogan, or mine, as a precise
characterization of everything in the practice of mathematics. Read
\emph{Mathematics, Form and Function} for a careful discussion of the
rich variety of activities involved in the actual practice of
mathematics.

The reflexivity of formal systems that we observed in regard to
Example~\ref{exa:SK} may easily lead the unwary from a view of
mathematics as a study heavily concerned with formal systems, back to
the ``vulgar formalist'' view of mathematics as a particular formal
system. If patterns of derivations in formal systems, and in classes
of formal systems, can be modeled successfully by other formal
systems, perhaps there is no real difference between the vulgar and
refined versions of formalism. But, the vulgar view leads to an
infinite regress of modeling systems in one another, explained very
nicely by Lewis Carroll in ``What the tortoise said to Achilles.'' As
a last defense, the vulgar formalist might point out that we may
choose an individual formal system, such as the Combinator Calculus,
that is capable of modeling the behavior of \emph{every} other formal
system. Perhaps this cuts off the infinite regress, by taking the
Combinator Calculus as the single source of all formal description (I
can't resist comparing it to the bottom turtle in the famous story of
the world being supported by ``turtles all the way down.''). No, there
is still an infinite regress of \emph{modeling} steps. The Combinator
Calculus provides a single language in which each of the modeling
steps may be described, but it does not cure the infinitude of steps.

\paragraph{Characterizing formal systems in general.}
I have avoided anything like a general definition of formal system
until now, because I think that it's hard to appreciate the components 
of such a definition without some previous observation of formal
systems and their uses.
\begin{definition}[Formal system]
\label{def:formsys}
\begin{itemize}
\item A \emph{formal alphabet} is a finite set of discrete symbols,
reliably distinguishable from one another.
\item A \emph{formal language} is a set of finite discrete arrangements
of symbols from a given formal alphabet, with a clear and unambiguous
characterization of the relevant qualities of an arrangement.
\item A \emph{formal system} is a system of rules for deriving
some of the arrangements of symbols from a given formal language, with 
a clear and unambiguous characterization of the manipulations that are 
and are not allowed as steps in such derivations.
\end{itemize}
\markendlst{End Definition~\ref{def:formsys}}
\end{definition}
Mathematical treatments of formal systems tend to use particular
conventional styles of presentation of formal alphabets, languages,
and rules, leading to the illusion that formality consists of
following such a style. Following accepted mathematical style is not
necessary for the presentation of a formal system---it is certainly
feasible to establish other conventions that are equally as clear and
unambiguous. Nor is it sufficient---it requires some
pre-indoctrination in the rules and meaning of the conventional
style. But, the use of mathematically conventional style is generally
efficient, since the pre-indoctrination is reusable for presenting
lots of different formal systems. What is crucial, though, is the
clear and unambiguous mutual understanding of all parties to a
discussion about a formal system. Often, that understanding is best
established by some questions and answers with examples. In practice,
we can be highly successful in establishing mutual understanding of
the rules of a formal system.

Finally, why do I present a study of formal systems as part of my
program to apply ideas from computer science to other disciplines? A
derivation in a formal system is precisely the same thing as a
computation. Computational systems and formal systems are the same
things, and studies of the two differ in the qualities that one is
motivated to consider, rather than the objects under study. Alan
Turing's ``On computable numbers, with an application to the
Entscheidungsproblem,'' and Emil Post's ``Finite combinatory
processes'' explore the human capacity for carrying out computation in
a pseudophysical style that complements nicely Curry's more
methodological study.

\section{The Origin of Formal Systems}
The concept of formal systems was not designed by a committee to
satisfy a research contract---it arose from centuries of work by
mathematicians and philosophers who were addressing individual
problems with formal components, rather than seeking to design a
general formal method, and it was characterized rigorously only in the
early twentieth century. Nonetheless, I propose that we understand
formal systems as a concept that is adapted very precisely to meet
sensible goals, even though its design was much more evolutionary than
conscious. The value of formal systems can best be appreciated by
considering how a fictitious conceptual engineer might have designed
them.

The earliest useful formal system that I can identify is the system of
nonnegative integers, also called the counting numbers. We are so
familiar with the counting numbers that many people sense that they
are real objective physical objects. I find that numbers are real and
objective, but not physical. The physical view of numbers appears to
be founded on experiences with collections of discrete objects, such
as pebbles placed in a bowl. Arguably $3+1=4$ is an observable
physical fact, because the result of placing $3$ pebbles in a bowl,
then adding $1$ more pebble, yields $4$ pebbles in the bowl. But, the
physical observations that support the concept of counting numbers
must be filtered by some previous understanding of number. When we
place $3$ drops of water in a bowl, then add $1$ more drop of water,
we find $1$ drop of water in the bowl. This experience does not lead
us to question the validity of $3+1=4$, nor even to view it as a mere
approximation of the truth about numbers. Rather, it leads us to
conclude that the number of distinguishable drops of water in a bowl
does not follow the rules of addition of counting numbers.

If the counting numbers are not physical observables, are they merely
psychological ephemera, or even illusions? I think not. The actual
practice of arithmetic in the world suggests that we understand the
counting numbers very well as essentially conceptual constructs in a
formal system (Curry called them \emph{formal objects}, or \emph{obs}
for short). A bit more carefully, the system of counting numbers
represents the common properties of a large class of formal systems,
each of which contains objects representing the numbers and
derivations corresponding to arithmetic calculations with
numbers. \mbox{$3+1=4$}, then, is an observation about the qualities
of the formal systems of counting numbers. We may convince ourselves
very reliably that $3+1=4$ by choosing a particular formal system, and
carrying out a derivation similar to the one in
Example~\ref{exa:incr}. In fact, the practice of placing pebbles in a
bowl can be understood as a presentation of a formal system for the
counting numbers---perhaps as an approximate presentation, since the
bowl has a limited capacity.

The usefulness of formal systems (and the effectiveness of
mathematics) derives from the fact that many natural phenomena are
exact or approximate presentations of formal systems. If we can
identify such formal systems and discover their properties, we can
characterize some of the properties of the corresponding natural
phenomena. Since formal systems may use any precisely and
unambiguously characterized notion of pattern, we have a chance to
apply formal systems to all natural phenomena in which we recognize
such patterns. Reasoning about formal systems is useful because it
often tells us that the presence of one pattern that we have observed
entails the presence of another pattern that we have not noticed. For
example, the pattern of definition of numerical addition in terms of
successor entails that the result of addition is independent of order.
This famous pattern is the \emph{commutative property} of addition,
written $x+y=y+x$ in algebraic texts.

Our understanding even of primitive propositions, such as $3+1=4$,
involves the recognition of patterns among formal systems, rather than
the mere exercise of a single formal system. Nobody really endorses
$3+1=4$ just because of the specific derivation in
Example~\ref{exa:incr}. Rather, we recognize that a large class of
formal systems exhibit certain qualities in common, which we regard as
the form of numerical arithmetic, and that all of these systems share
a pattern which we describe by $3+1=4$. The derivation in
Example~\ref{exa:incr} is simultaneously an example of the $3+1=4$
pattern in a particular formal system, and a formal presentation of
the pattern of reasoning that we follow in recognizing the $3+1=4$
pattern as an unavoidable consequence of the qualities shared by all
of the formal presentations of numerical arithmetic.

If they are mental and abstract, rather than physical, how can formal
systems be real and objective? They were evolved to be so, by
eliminating from mental processes all the parts that we cannot agree
on reliably. Suppose that we tasked our fictitious conceptual engineer
to produce a method of reasoning, and communicating the results of
reasoning, with the greatest possible capability for
\begin{itemize}
\item minimizing the physical cost of reasoning and communication, and
\item maximizing objective certainty in the conclusions derived by
reasoning.
\end{itemize}
If he were sufficiently brilliant, our engineer might consider that
manipulation of symbols can be made physically very cheap, because we
are always at liberty to substitute a lighter weight symbol for a
heavier one, as long as all parties to a discussion recognize the same 
selection of symbols. Next, the rules for manipulating symbols should
be based on the forms of arrays of those symbols, rather than a
particular assigned meaning, in order to allow reasoners to maximize
objective certainty. If the rules for manipulating symbols depend on
their meanings, then our certainty about the correctness of
manipulations is limited by our certainty about the behavior of the
things that they refer to. We would like to use reasoning to
illuminate qualities of things about which we are initially quite
uncertain. So, we choose rules that refer only to the forms of arrays
of symbols, and then we are at liberty to present the arrays and the
rules in ways that we have discovered in practice to be thoroughly
clear and unambiguous. It appears that our conceptual engineer has
just designed for us the highly successful method of reasoning with
formal systems.

Those with a taste for ontological studies may find my treatment of
formal systems disturbingly circular. Instead of characterizing the
essence of formal systems, and showing why that essence leads to
certainty about the results of reasoning, I suggest that formal
systems are the systems that we design for ourselves by refusing to
deal with material about which we are uncertain. I concede the
circularity, but I think that it represents the best way of
understanding formal systems. They are social objects, designed for
the purpose of communication. Because the design is so successful,
they acquire an a posteriori air of necessity. In the next section, I
trace the quest for certainty through the work of Descartes and
Hilbert, and I think that this circular and somewhat negative view is
at least highly consistent with their published ideas.

Another famously successful early example of a formal system is
Euclidean geometry. Practitioners of geometry did not show clear
understanding of the formal quality that characterized their work, but
they had a strong intuitive sense that this work was reliable in a way
quite different from other philosophical inquiries. Oddly, geometry
was held for centuries to be an exact description of the layout of the 
physical universe, and the coexistence of formal certainty with
physical factuality puzzled thinkers, such as Emmanuel Kant. Now we
understand very well the sense in which geometry is a formal system
and the resulting certainty in its conclusions, but we no longer
believe that it describes the physical universe exactly.

\section{Descartes' and Hilbert's Quests for\\ Certainty}
\paragraph{Descartes' \emph{Discourse} as a task description for the
design of formal systems.}
In his \emph{Discourse on the Method of Rightly Conducting the Reason,
and Seeking Truth in the Sciences}, Ren\'e Descartes sought ``clear
and certain knowledge of all that is useful in life.'' He described a
method that he expected to lead to such ``clear and certain
knowledge,'' given enough time and careful effort. Although it does
not explicitly distinguish formal from contentual reasoning, the
\emph{Discourse} may be understood as containing a task description
for our fictitious engineer, and substantial good advice to lead her
toward her brilliant design.

Although he never showed an explicit awareness of something like
formal systems, Descartes recognized mathematics, and geometry in
particular, as a domain in which certainty had been achieved.
\begin{quote}
I was especially delighted with the mathematics, on account of the
certitude and evidence of their reasonings; but I had not as yet a
precise knowledge of their true use; and thinking that they but
contributed to the advancement of the mechanical arts, I was
astonished that foundations, so strong and solid, should have had no
loftier superstructure reared on them. On the other hand, I compared
the disquisitions of the ancient Moralists to very towering and
magnificent palaces with no better foundation than sand and mud: they
laud the virtues very highly, and exhibit them as estimable far above
anything on earth; but they give us no adequate criterion of virtue,
and frequently that which they designate with so fine a name is but
apathy, or pride, or despair, or parricide.
\end{quote}
In the description of his method, Descartes does not explicitly
mention the need for objective judgments, nor the need to communicate 
reasoning to others. But, in the supporting material he claims that
others can follow his own method, and in particular that even children 
may follow the rules of arithmetic.
\begin{quote}
The child, for example, who has been instructed in the elements of
Arithmetic, and has made a particular addition, according to rule, may
be assured that he has found, with respect to the sum of the numbers
before him, all that in this instance is within the reach of human
genius. Now, in conclusion, the Method which teaches adherence to the
true order, and an exact enumeration of all the conditions of the
thing sought, includes all that gives certitude to the rules of
Arithmetic.
\end{quote}
Here Descartes mentions his admiration of geometry.
\begin{quote}
The long chains of simple and easy reasonings by means of which
geometers are accustomed to reach the conclusions of their most
difficult demonstrations, had led me to imagine that all things, to
the knowledge of which man is competent, are mutually connected in the
same way, and that there is nothing so far removed from us as to be
beyond our reach, or so hidden that we cannot discover it, provided
only we abstain from accepting the false for the true, and always
preserve in our thoughts the order necessary for the deduction of one
truth from another.
\end{quote}
And here he claims that the conclusions of geometry are certain
precisely because the method of geometry follows his rules.
\begin{quote}
I was disposed straightway to search for other truths; and when I had
represented to myself the object of the geometers, which I conceived
to be a continuous body, or a space indefinitely extended in length,
breadth, and height or depth, divisible into divers parts which admit
of different figures and sizes, and of being moved or transposed in
all manner of ways, (for all this the geometers suppose to be in the
object they contemplate,) I went over some of their simplest
demonstrations. And, in the first place, I observed, that the great
certitude which by common consent is accorded to these demonstrations,
is founded solely upon this, that they are clearly conceived in
accordance with the rules I have already laid down.
\end{quote}
Although Descartes probably did not understand completely the sense in
which geometry is a formal system, it is telling that he picks a
formal system as the only clear example of the success of his own
method. It is reasonable to conclude that Descartes intends his
method, at least when it is applied to mathematical topics, to
coincide with the method of reasoning with formal systems.

I must concede that Descartes' conception of the certainty of geometry
was not the same as the modern one. He surely believed that the
postulates of Euclidean geometry represent certain and axiomatic
knowledge about the physical universe. Relativistic physics does not
support the correctness, much less the certainty, of the parallel
postulate, and quantum physics suggests fundamental flaws in the
concepts through which the postulates are expressed. The certainty in
geometry now appears only to be that the postulates entail a rich set
of additional information about Euclidean configurations. I prefer to
imagine that Descartes had a correct, but vague, intuition about the
real certainty in geometric deductions, and was wrong only about the
certainty of its postulates, rather than to imagine that he had
something radically different from a formal system of geometry in
mind.

Now, look at Descartes' description of his method. This is important
enough to quote in full, and in several versions. I give, first, the
English translation by John Veitch, then the original French written
by Descartes, then the Latin translation by Descartes' friend Etienne
De Courcelles, which was revised by Descartes himself, and finally an
English translation by Laurence J.\ Lafleur that combines the French
version with the Latin: portions in parentheses are from the Latin
only, in square brackets French only, in other portions the French and
Latin agree.

Descartes' first rule states the requirement of certainty in the
results of reasoning, although it does not mention objectivity or
agreement between different reasoners. Its style is consistent with my
suggestion that we achieve certainty by rejecting all material that is
unclear or ambiguous.
\begin{quote}
The \emph{first} was never to accept anything for true which I did not
clearly know to be such; that is to say, carefully to avoid
precipitancy and prejudice, and to comprise nothing more in my
judgement than what was presented to my mind so clearly and distinctly
as to exclude all ground of doubt.

Le premier était de ne recevoir jamais aucune chose pour vraie que je
ne la connusse évidemment être telle: c'est-à-dire d'éviter
soigneusement la précipitation et la prévention; et de ne comprendre
rien de plus en mes jugements, que ce qui se présenterait si
clairement et si distinctement à mon esprit, que je n'eusse aucune
occasion de le mettre en doute.

Primum erat, ut nihil unquam veluti verum admitterem nisi quod cert\`o 
\& evidenter verum esse cognoscerem; hoc est, ut omnem pr{\ae}cipitantiam
atque anticipationem in judicando diligentissim\`e vitarem; nihilque
amplius conclusione complecterer qu\`am quod tam clar\`e \&
distinct\`e rationi me{\ae} pateret, ut nullo modo in dubium possem
revocare.

The first rule was never to accept anything as true unless I
recognized it to be (certainly and) evidently such: that is, carefully
to avoid (all) precipitation and prejudgment, and to include nothing
in my conclusions unless it presented itself so clearly and distinctly
to my mind that there was no (reason [or) occasion] to doubt it. The
second, to divide each of the difficulties under examination into as
many parts as possible, and as might be necessary for its adequate
solution.
\end{quote}

The second rule does not connect directly to my discussion above, but
it resembles the reasoning used by Turing and Post in their
computational versions of formal systems, where they divided
computational steps into pieces small enough that there could be no
doubt about their correctness.
\begin{quote}
The \emph{second}, to divide each of the difficulties under
examination into as many parts as possible, and as might be necessary
for its adequate solution.

Le second, de diviser chacune des difficultés que j'examinerais, en
autant de parcelles qu'il se pourrait et qu'il serait requis pour les
mieux résoudre.

Alterum, ut difficultates, quas essem examinaturus, in tot partes
dividerem, quot expediret ad illas commodi\`us resolvendas.

The second was to divide each of the difficulties which I encountered
into as many parts as possible, and as might be required for an easier
solution.
\end{quote}

The third rule does not make a definitive connection to formal
systems, but it is certainly consistent with the progression, in
formal systems of logical reasoning, from simple postulates to more
subtle and complex theorems.
\begin{quote}
The \emph{third}, to conduct my thoughts in such order that, by
commencing with objects the simplest and easiest to know, I might
ascend by little and little, and, as it were, step by step, to the
knowledge of the more complex; assigning in thought a certain order
even to those objects which in their own nature do not stand in a
relation of antecedence and sequence.

Le troisième, de conduire par ordre mes pensées, en commençant par les
objets les plus simples et les plus aisés à connaître, pour monter peu
à peu, comme par degrés jusqu'à la connaissance des plus composés, et
supposant même de l'ordre entre ceux qui ne se préc\`edent point
naturellement les uns les autres.

Tertium ut cogitationes omnes, quas veritati qu{\ae}rend{\ae}
impenderem, certo semper ordine promoverem: incipiendo scilicet \`a
rebus simplicissimis \& cognitu facillimis, ut paulatim, \& quasi per
gradus, ad difficiliorum \& magis compositarum cognitionem ascenderem; 
in aliquem etiam ordinem illas mente disponendo, qu{\ae} se mutu\`o ex 
natura sua non pr{\ae}cedunt.

The third was to think in an orderly fashion (when concerned with the
search for truth), beginning with the things which were simplest and
easiest to understand, and gradually and by degrees reaching toward
more complex knowledge, even treating, as though ordered, materials
which were not necessarily so.
\end{quote}

The fourth rule also expresses a necessary, but not definitive,
quality of formal systems.
\begin{quote}
And the \emph{last}, in every case to make enumerations so complete,
and reviews so general, that I might be assured that nothing was
omitted.

Et le dernier, de faire partout des dénombrements si entiers, et des
revues si générales que je fusse assuré de ne rien omettre.

Ac postremum, ut tum in qu{\ae}rendis mediis, tum in difficultatum
partibus percurrendis, tam perfect\`e singula enumerarem \& ad omnia
circumspicerem, ut nihil \`a me omitti essem certus.

The last was (, both in the process of searching and in reviewing when
in difficulty,) always to make enumerations so complete, and reviews
so general, that I would be certain that nothing was omitted.
\end{quote}

One cannot discuss the \emph{Discourse} without mentioning its most
famous sentence: ``Je pense, donc je suis,'' ``Cogito ergo sum,'' ``I
think, therefore I am.'' I regard this fascinating sentence as an
unsuccessful attempt to apply the four rules. In particular, it
represents an attempt to provide a new postulate about existence with
the same sort of certainty that Descartes associated, incorrectly,
with the postulates of geometry. While formal systems provide very
strong certainty about their derivations, his proposed extension of
certain knowledge to fundamental postulates is probably impossible to
achieve.

\paragraph{Hilbert restricted Descartes' program to mathematics.}
In ``The foundations of mathematics,'' and ``On the infinite,'' David
Hilbert proposed a program to ``recast mathematical definitions and
inferences in such a way that they are unshakable.'' Hilbert relied
explicitly on formal systems as the tool for achieving unshakable
certainty regarding all of mathematics. Although he did not refer
explicitly to Descartes, we may read the first part of Hilbert's
program as the restriction of Descartes' program to mathematics,
instead of ``all that is useful in life.''

In these lectures, Hilbert repeatedly uses the German word
``inhaltlich.'' The word has no precise English counterpart in common
usage. I have taken Stefan Bauer-Mengelberg's translation to
``contentual,'' which I understand to mean \emph{referring to the
content or meaning of assertions or formulae}. ``Contentual'' is a
sensible opposite to ``formal.'' Curry translated ``inhaltlich'' as
``contensive.'' It is sometimes rendered as ``intuitive,'' which I
find quite misleading. It seems clear that Hilbert is distinguishing
between form and content---one may employ intuition in dealing with
either of these.  He refers at least once to the application of
``perceptual intuition'' to forms.

Hilbert's program has two major steps:
\begin{enumerate}
\item to recast all of mathematics within one formal system of
reasoning, and
\item to prove, using only elementary reasoning about finite objects,
the consistency of that formal system.
\end{enumerate}
Step one may be understood as the restriction of Descartes' program to
mathematics. If step one were successful, then we would have a uniform
formal mechanism for deriving all mathematical truths, but we would
still take time to produce each individual truth. We would enjoy
thorough certainty that our derivations were correct in terms of the
rules of the formal system, but not that the rules that themselves
were appropriate. Step two gives primacy to the generation of one
particular truth intended to secure the correctness of step one
against every conceivable challenge. Hilbert was reacting in
particular to the discovery of contradictions in basic set theory and
the study of infinitesimals in the differential and integral calculus,
which he considered as mathematical catastrophes. Although step two
holds the key to Hilbert's motivation, I am concerned here with step
one.

Hilbert expresses his intention to achieve certainty through formal
systems quite explicitly.
\begin{quote}
I should like to eliminate once and for all the questions regarding
the foundations of mathematics, in the form in which they are now
posed, by turning every mathematical proposition into a formula that
can be concretely exhibited and strictly derived, thus recasting
mathematical definitions and inferences in such a way that they are
unshakable and yet provide an adequate picture of the whole science.
\end{quote}

\begin{quote}
In my theory contentual inference is replaced by manipulation of signs 
according to rules; in this way the axiomatic method attains that
reliability and perfection that it can and must reach if it is to
become the basic instrument of all theoretical research.
\end{quote}
\begin{quote}
[My theory's] aim is to endow mathematical method with \dots definitive
reliability.
\end{quote}
\begin{quote}
It is necessary to formalize the logical operations and also the
mathematical proofs themselves.
\end{quote}
\begin{quote}
A formalized proof, like a numeral, is a concrete and surveyable object.
\end{quote}

In mathematical practice, concepts of infinite objects are the ones
that seem to call on questionable contentual intuitions, so Hilbert is 
particularly concerned with formalizing those concepts.
\begin{quote}
Modes of inference employing the infinite must be replaced generally
by finite processes that have precisely the same results.
\end{quote}

Although Hilbert insists on formal systems as the only sources of
certainty in mathematical inference, he appeals to direct contentual
intuition for the truth of basic numerical equations, such as
$3+1=4$.
\begin{quote}
We recognize that we can obtain and prove [numerical] truths through
contentual intuitive considerations.
\end{quote}
I claim that our certainty about integer arithmetic derives from the
essentially formal nature of the numbers (that is, the content of
integer arithmetic is a particular sort of form). Hilbert insists on a
distinction between the contentual appreciation of integer arithmetic
and the formal description of other concepts in mathematics (he calls
them ``ideal'' concepts, to indicate that they are not significant in
themselves, but only as ways of organizing arithmetic truths). But,
the certainty of integer arithmetic derives from our ability to check
each formula by computation, which is just derivation in a formal
system. I think it is fair to say that both the arithmetic ground of
Hilbert's mathematics and its ``ideal'' superstructure are essentially
formal systems, but that he considers the formal systems of arithmetic
to be uniquely chosen for some prior reasons, while he regards the
rules for reasoning about ideals as products of a less constrained
design. In my view, the content of integer arithmetic is a sort of
form. Hilbert chooses to emphasize its status as content, largely to
combat the criticism of mathematics as merely a game played with
symbols. But, our certainty about integer arithmetic derives from its
formal nature.

Hilbert treats the concept of formal systems as a pre-existing tool to 
be used in his work. He mentions some key qualities of formal systems, 
but he does not inquire explicitly into the origin of formal systems
and the sources of certainty.
\begin{quote}
If logical inference is to be reliable, it must be possible to
survey [mathematical] objects completely in all their parts, and the
fact that they occur, that they differ from one another, and that they 
follow each other, or are concatenated, is immediately given
intuitively, together with the objects, as something that neither can
be reduced to anything else nor requires reduction.
\end{quote}
Some passages at least suggest that we may regard formal systems as
the natural outcome of design requirements. The phrase below,
``according to the conception we have adopted,'' seems to acknowledge
that formalisms are objective partly because we adopt precisely those
formal distinctions that all parties agree to recognize.
\begin{quote}
In mathematics, in particular, what we consider is the concrete signs
themselves, whose shape, according to the conception we have adopted,
is immediately clear and recognizable.
\end{quote}
Hilbert's comparison of his approach to the foundations of mathematics 
to other proposals is consistent with the idea of formal systems as
engineered concepts.
\begin{quote}
Mathematics is a presuppositionless science. To found it I do not need
God, as does Kronecker, or the assumption of a special faculty of our
understanding attuned to the principle of mathematical induction, as
does Poincar\'e, or the primal intuition of Brouwer, or, finally, as
do Russell and Whitehead, axioms of infinity, reducibility, or
completeness, which in fact are actual, contentual assumption that
cannot be compensated for by consistency proofs.
\end{quote}
A final passage acknowledges the importance of universal agreement,
although it does not explain how formal systems lead to such
agreement.
\begin{quote}
Mathematics in a certain sense develops into a tribunal of
arbitration, a supreme court that will decide questions of
principle---and on such a concrete basis that universal agreement must 
be attainable and all assertions can be verified.
\end{quote}

\section{The Strength and Scope of Formal\\ Certainty}
\paragraph{How certain are formal derivations?} Not \emph{absolutely}
certain, since they depend on consensus regarding formal distinctions,
and the correct perception of those formal distinctions through
whatever senses we choose for their presentation. I find this not very
disturbing, and doubt the possibility of absolute certainty about
anything. The correctness of formal derivations is at least as certain
as primitive sensual observations, such as \emph{the sky is blue} and
\emph{the sun rose this morning.} They are more robust, since we are
at liberty to repeat the verification of a derivation to the limits of 
our attention and tolerance for tedium, and to recast the presentation
of symbols whenever we notice a potential for error or
ambiguity. Although not absolute, formal derivations arguably enjoy
the highest degree of objective certainty that is attainable by
rational intelligence.

The strength of certainty is not purely quantitative---there are
different sorts of certainty. When we stand on Gibraltar, we feel an a
priori sort of certainty in the independent quality of that rock as a
support for our feet. Certainty in the derivation of formal systems is
a more social sort of certainty. It shares some of the qualities of
our certainty in an automobile that is warranted by a reliable
firm. We are confident, but not that the auto will always function
perfectly. Rather, we are confident that we can recognize deviations,
and adjust the machine, with occasional appeal to the maker, so that
it eventually gets us where we want to go. Similarly, we are certain
about derivations in formal systems because we can detect errors, and
we can refine our physical presentations of arrays of symbols to
overcome momentary confusion and ambiguity. Because of the extreme
efficiency and malleability of the basic symbolic materials underlying 
formal systems, the degree of certainty in final success is much
stronger than the degree of certainty in even the best engineered
automobiles.

\paragraph{What assertions are we so certain about?} Strictly, a
formal system only gives us strong certainty that certain derivations
do or do not follow the rules of the system. They do not and cannot
provide certainty that particular natural phenomena, such as the
configuration of paths followed by particles of light, follow
precisely the rules of a formal system, such as Euclidean or
non-Euclidean geometry.  But, the scope of formally derived certainty
is much more valuable practically than this mere certainty relative to
the rules suggests to pessimists. Reasoning about formal systems can
give us extremely high certainty that the presence of one formal
pattern entails the presence of another. Since our observations of the
universe abound in, and arguably consist entirely of, recognitions of
formal patterns, the actual effectiveness of formal systems is
substantial, and not at all unreasonable.

\paragraph{What are the formal limits on formal studies?} G\"odel's
famous incompleteness theorem shows the first step in Hilbert's
program to be inherently impossible to achieve. No single formal
system can derive all of the truths of integer number theory. If we
accept that Descartes' program contains the first step in Hilbert's,
then his program is also inherently impossible. The second step in
Hilbert's program depends on the first. But, if we choose a single
formal system that is sufficient for some part of mathematical
practice, there is still value in proving the consistency of that
limited system. G\"odel also showed that consistency of one formal
system requires reasoning that is in some technical sense too powerful
to be carried out in the system under study. It is natural to conclude
that Hilbert's second step is impossible, even accepting a limited
accomplishment of the first step. Takeuti pointed out that the
technical power of a system involved in G\"odel's theorem is not
necessarily connected to the ontological level of our confidence in
the system. So, it makes sense to prove the consistency of a formal
system using rules of reasoning that are technically more powerful,
but intuitively more secure, than those of the system under
investigation. The practical impact of this approach to Hilbert's
second step has only been partially explored.

\section{More to Think About}
\begin{itemize}
\item For a deeper look at many of the issues introduced in this
article, here are some further readings from the bibliography.
\begin{itemize}
\item Haskell B.\ Curry explores formalism as the content of
mathematics in \emph{Outlines of a Formalist Philosophy of
Mathematics} and \emph{Combinatory Logic, Volumes I and II}.  The two
books on \emph{Combinatory Logic} are mostly full of mathematical
technicalities, but the first chapter of each discusses the philosophy
of formalism and its relation to mathematics. In particular,
Chapter~11, the first of Volume II, reacts to misunderstanding of the
presentation in Chapter 1 of Volume I, perhaps by the same vulgar
formalists who annoyed Mac Lane.
\item A.\ M.\ Turing's ``On computable numbers'' and E.\ Post's
``Finite combinatory processes'' explore the way that the structure of
computation derives from the physical processes involved in computing
by humans.
\item R.\ W.\ Hamming  and E.\ P.\ Wigner, in ``The unreasonable
effectiveness of mathematics \dots,'' explore the practice of pure and 
applied mathematics, and describe some phenomena that support the
notion of mathematics as the result of conceptual engineering.
\item S.\ Mac Lane, in \emph{Mathematics, form and function}, gives
the most thorough and accurate treatment that I have seen of the rich
entanglement of mathematics with formal systems at a number of levels.
\item I annotated Descartes' \emph{Discourse}, and
the two lectures by Hilbert, in somewhat more detail for a college
course at the University of Iowa. You may view the annotations on the
World Wide Web at \linebreak
\emph{http://www.cs.uchicago.edu/\~\relax  odonnell/OData/Courses/22C:096/\linebreak Lecture\_notes/contents.html}.
\end{itemize}
\item Investigate the actual evolution of systems of symbols, using
linguistic and psychological methods to illuminate the weeding out of
ambiguity in recognizing symbols and their arrangements (not the same
as ambiguity in their meanings). Investigate the interaction with
efficiency of presentation.
\item Find examples for early successful uses of formal systems,
besides integer arithmetic and geometry. Seek, in particular, more
primitive systems that may not have been associated consciously with
mathematics.
\item Find examples for natural phenomena with formal properties as
reliable as the numerical properties of pebbles in a bowl. Notice
limits on the exactness of even the best of these formal descriptions
of nature (for example, limits on the number of pebbles that might
ever be contained in a bowl).
\item Trace the changing attitude toward formal geometry over the centuries.
\item Follow the precursors of, and explicit references to, computation and 
formal systems through the works of other philosophers, particularly
Emmanuel Kant.
\item Investigate the importance of the reflexivity of formal
systems---the formal study of formalism. Does it contribute to the
certainty achieved by such systems? How does it affect the character
of formal studies?
\item Present the story of Hilbert's program as a tragedy in the
formal sense defined by Aristotle.
\item Analyze computational systems as systems designed to have
objectively communicable results. Many qualities of computational
systems may derive from the limitations of robust communication, due
to the information bottleneck imposed by language. There might be a
new type of evidence for the Church-Turing thesis here (thesis: Turing
machines, Combinatory Calculus, and some other known formal systems
characterize precisely the computable functions).
\end{itemize}

\nocite{carroll, curry-formalist, curry-combinator-1, curry-combinator-2,
discourse-vietch, discourse-lafleur, discours, discours-gadoffre,
hamming, hilbert-infinite, hilbert-foundations,
hilbert-infinite-heijenoort, hilbert-foundations-heijenoort, maclane,
post-compute, post-compute-davis, post-critique, post-critique-davis,
takeuti, turing-computable, turing-computable-davis, turing-ace,
turing-ace-mit, wigner}

\bibliographystyle{abbrv}

\bibliography{certainty}

\end{document}